\documentclass[pra,aps,twocolumn,epsf,amsfonts]{revtex4}

\usepackage{graphicx}% Include figure files
\usepackage{epsfig}
\def\qed{\leavevmode\unskip\penalty9999 \hbox{}\nobreak\hfill
     \quad\hbox{\leavevmode  \hbox to.77778em{%
               \hfil\vrule   \vbox to.675em%
               {\hrule width.6em\vfil\hrule}\vrule\hfil}}
     \par\vskip3pt}

\def\ibb #1{\leavevmode\hbox{\kern.3em\vrule
     height 1.5ex depth -.1ex width .4pt\kern-.3em\rm#1}}

\newcommand{\bea}[1]{\begin{eqnarray} #1 \end{eqnarray} }
\newcommand{\ba}[2]{\begin{array}{#1}#2\end{array}}
\newcommand{\tr}[1]{{\rm tr}\left[#1\right]}

\begin{document}

\title{Activating NPPT distillation with an infinitesimal amount of bound entanglement}

\author{Karl Gerd H. Vollbrecht}
\email{k.vollbrecht@tu-bs.de}
\author{Michael M. Wolf}
\email{mm.wolf@tu-bs.de} \affiliation{Institute for Mathematical
Physics, TU Braunschweig, Germany}

\date{\today}

\begin{abstract}
We show that bipartite quantum states of any dimension, which do
not have a positive partial transpose, become 1-distillable when
one adds an infinitesimal amount of bound entanglement. To this
end we investigate the activation properties of a new class of
symmetric bound entangled states of full rank. It is shown that in
this set there exist universal activator states capable of
activating the distillation of any NPPT state.
\end{abstract}


\maketitle

\section{Introduction}

The possibility of {\it distillation} plays a crucial role in
Quantum communication and Quantum information processing (cf.
\cite{Springer}). Together with quantum error correction it
enables all the fascinating applications provided by Quantum
information theory in the presence of a noisy and interacting
environment. Despite its practical relevance and quite
considerable effort in that direction, however, many of the basic
questions concerning distillation are yet unanswered. Most notably
the question whether a given quantum state is distillable or not,
i.e., whether it is possible to obtain pure maximally entangled
states from several copies of it by means of local operations and
classical communication (LOCC).

A necessary condition for the distillability  of a state described
by a density matrix $\rho$ is the fact that its partial transpose
$\rho^{T_A}$, defined with respect to a given product basis by
$\langle ij|\rho^{T_A}|kl\rangle=\langle kj|\rho| il\rangle$, has
a negative eigenvalue \cite{HHH2}. Except for special cases like
states on ${\mathbb{C}}^2\otimes{\mathbb{C}}^n$ \cite{2n,IHa} and
Gaussian states \cite{Gauss} it is however unclear whether this
condition is sufficient as well. There is some evidence presented
in \cite{IHa,IBM} that this may not be the case and that there are
indeed undistillable states, whose partial transpose is not
positive (NPPT). At least there exist $n$-undistillable NPPT
states for every finite $n$, meaning that no LOCC operation on $n$
copies leads even to a single entangled two qubit state
\cite{IHa,IBM}.

However, if we enlarge the class of allowed distillation protocols
from LOCC to channels respecting the positivity of the partial
transpose, then every NPPT state becomes 1-distillable \cite{EVWW,
KLC2} (which can be shown by using {\it entanglement witnesses}
\cite{KLC2}). Moreover, it is a result from \cite{CDK} that these
channels can always be stochastically implemented by an LOCC
operation where the two parties are given an entangled state with
positive partial transpose (PPT) as an additional resource. The
latter is known to be {\it bound entangled} since the entanglement
needed for the preparation of the state cannot be recovered by
distillation \cite{HHH2}. Nevertheless, PPT bound entangled states
can be useful in order to {\it activate} the distillability of
bipartite NPPT states \cite{Terhal, KLC}.

The aim of the present paper is to investigate the limits and
requirements of such an activation process. We will show that
there exist states with an arbitrary small amount of PPT bound
entanglement, which are capable of activating any NPPT state. The
required additional resource is therefore universal as well as
arbitrary weakly entangled.

\section{Preliminaries on symmetric states}

One of the key ideas in what follows will be the exploitation of
the symmetry properties of states commuting with certain local
unitaries. Two well known one-parameter families of such states
are the {\it Werner states} and {\it Isotropic states}, both
playing an important role in the sequel.

{\it Werner states} \cite{W89} acting on a Hilbert space ${\cal
H}={\cal H}_A\otimes{\cal H}_B$ with dimensions $\dim{\cal H}_A =
\dim{\cal H}_B = d$ commute with all unitaries of the form
$U\otimes U$ and can be written as
\begin{equation}\label{Werner}
\rho(\alpha)=\Big({\bf 1}-\frac{\alpha}d
\mathbb{F}\Big)/(d^2-\alpha) , \quad \alpha\in [-d,d] ,
\end{equation}
with $\mathbb{F}$ being the {\it flip operator} defined with
respect to some product basis by $\mathbb{F}|ij\rangle
=|ji\rangle$. A Werner state is entangled iff $\alpha\in (1,d]$
and 1-distillable iff $\alpha\in (d/2,d]$. Moreover, it was shown
in \cite{Iso} that any NPPT state can be mapped onto an entangled
Werner state by means of LOCC operations. Therefore we can in the
following restrict our discussion to the activation of Werner
states keeping in mind that the obtained results hold for any NPPT
state.

{\it Isotropic states} \cite{Iso,Sym} commuting with all unitaries
of the form $U\otimes\bar{U}$ (where $\bar{U}$ is the complex
conjugate of $U$) are combinations of the maximally mixed state
${\bf 1}/d^2$ and the projector
$\mathbb{P}=|\Omega\rangle\langle\Omega|$ onto the maximally
entangled state $|\Omega\rangle =1/\sqrt{d}\sum_{i=1}^d
|ii\rangle$:
\begin{equation}\label{Iso}
\omega(f)= f\mathbb{P} + \frac{1-f}{d^2-1} \big({\bf
1}-\mathbb{P}\big),\quad f\in [0,1] .
\end{equation}
 An Isotropic state $\omega$ is known to be 1-distillable iff the {\it maximally entangled fraction}
$f=\langle\Omega|\omega|\Omega\rangle
> 1/d$, which is a sufficient condition for any other state as
well \cite{Iso}. Hence, an activation protocol succeeded if this
condition is fulfilled by the final state.

The symmetric states playing the central role in the present paper
act on a larger Hilbert space ${\cal H}={\cal H}_{A_1}\otimes{\cal
H}_{A_2}\otimes{\cal H}_{B_1}\otimes{\cal H}_{B_2}$ of total
dimension $d^4$, where $A$ and $B$ again label the two parts of
the system situated at different locations. The symmetry group
under consideration is the group of all unitaries of the form
$W=(U\otimes V)_A\otimes(U\otimes\bar{V})_B$. States $\sigma$
commuting with all these unitaries can most easily be expressed in
terms of the minimal projectors $\{P_i\}$ spanning the commutant
of the group \cite{commutant}: \bea{
  \forall W : [\sigma, W]=0 \quad & \Leftrightarrow & \quad \sigma =
  \sum_{i=1}^4
  \lambda_i P_i / \tr{P_i},\label{UUUA}\\
 \mbox{with}\qquad\quad P_{{1}\atop{2}} &=& \frac12 \big({\bf 1}\mp \mathbb{F}\big)_1\otimes \mathbb{P}_2 , \\
P_{{3}\atop{4}} &=& \frac12 \big({\bf 1}\mp
\mathbb{F}\big)_1\otimes \big({\bf 1}-\mathbb{P}\big)_2 .} Note
that, as labeled by the indices, the tensor products correspond to
a split $1|2$ (and not $A|B$). Positivity and normalization of
$\sigma$ requires $\lambda_i \geq 0$ and $\sum_{i=1}^4
\lambda_i=1$ such that any state of the considered symmetry can be
characterized by a vector $\vec{\lambda}\in \mathbb{R}^3$ lying in
a tetrahedron, which is given by these constraints. We note
further that the set of symmetric states in (\ref{UUUA}) is
abelian, i.e., all symmetric states commute with each other.

The {\it activation protocol} we use follows closely an idea of
Ref. \cite{CDK}. Initially the two parties $A$ and $B$ are
supposed to share a Werner state $\rho(\alpha)$ acting on ${\cal
H}_0={\cal H}_{A_0}\otimes{\cal H}_{B_0}$ with $\dim{\cal H}_0 =
d^2$ and a symmetric state $\sigma$ on ${\cal H}_1\otimes{\cal
H}_2$ given by Eq.(\ref{UUUA}). After a local filtering operation
is applied by projecting onto maximally entangled states
$\mathbb{P}_{A_{0,1}}$ and $\mathbb{P}_{B_{0,1}}$ (acting on
${\cal H}_{A_0}\otimes{\cal H}_{A_1}$ respectively ${\cal
H}_{B_0}\otimes{\cal H}_{B_1}$) the maximally entangled fraction
of the resulting state on system 2 is given by
\begin{equation}\label{fsr}
f\big(\rho(\alpha);\sigma\big):=\frac{\tr{(\rho(\alpha)\otimes\sigma)(\mathbb{P}_{A_{0,1}}
\otimes\mathbb{P}_{B_{0,1}}
\otimes\mathbb{P}_2)}}{\tr{(\rho(\alpha)\otimes\sigma)
(\mathbb{P}_{A_{0,1}}\otimes\mathbb{P}_{B_{0,1}}\otimes{\bf 1}
_2)}}.
\end{equation}
We know that $\sigma$ {\it activates} $\rho(\alpha)$ if
$f\big(\rho(\alpha);\sigma\big)>1/d$. Since the output state of
the protocol is itself isotropic, this condition is also necessary
for the activation.

Of course, we are only interested in cases where $\alpha\in
(1,d/2]$, i.e., $\rho(\alpha)$ is entangled but not 1-distillable,
and $\sigma$ is in turn a PPT bound entangled state. The latter
requires the classification of the symmetric states in
(\ref{UUUA}), which is the content of the next section.

\section{Classification and Activation}

The following discussion will mainly take place in the three
dimensional space given by the expansion coefficients
$\vec{\lambda}=(\lambda_1,\lambda_2,\lambda_3)$ from Eq.
(\ref{UUUA}). A vector $\vec{\lambda}$ corresponds to a (positive
and normalized) symmetric state $\sigma(\vec{\lambda})$ iff
$\vec{\lambda}\in {\cal S}=\{ \vec{v}\in\mathbb{R}^3\ | v_i \geq
0, \sum_i v_i \leq 1\}$, where the state space ${\cal S}$ is a
tetrahedron.

\begin{figure}\epsfig{file=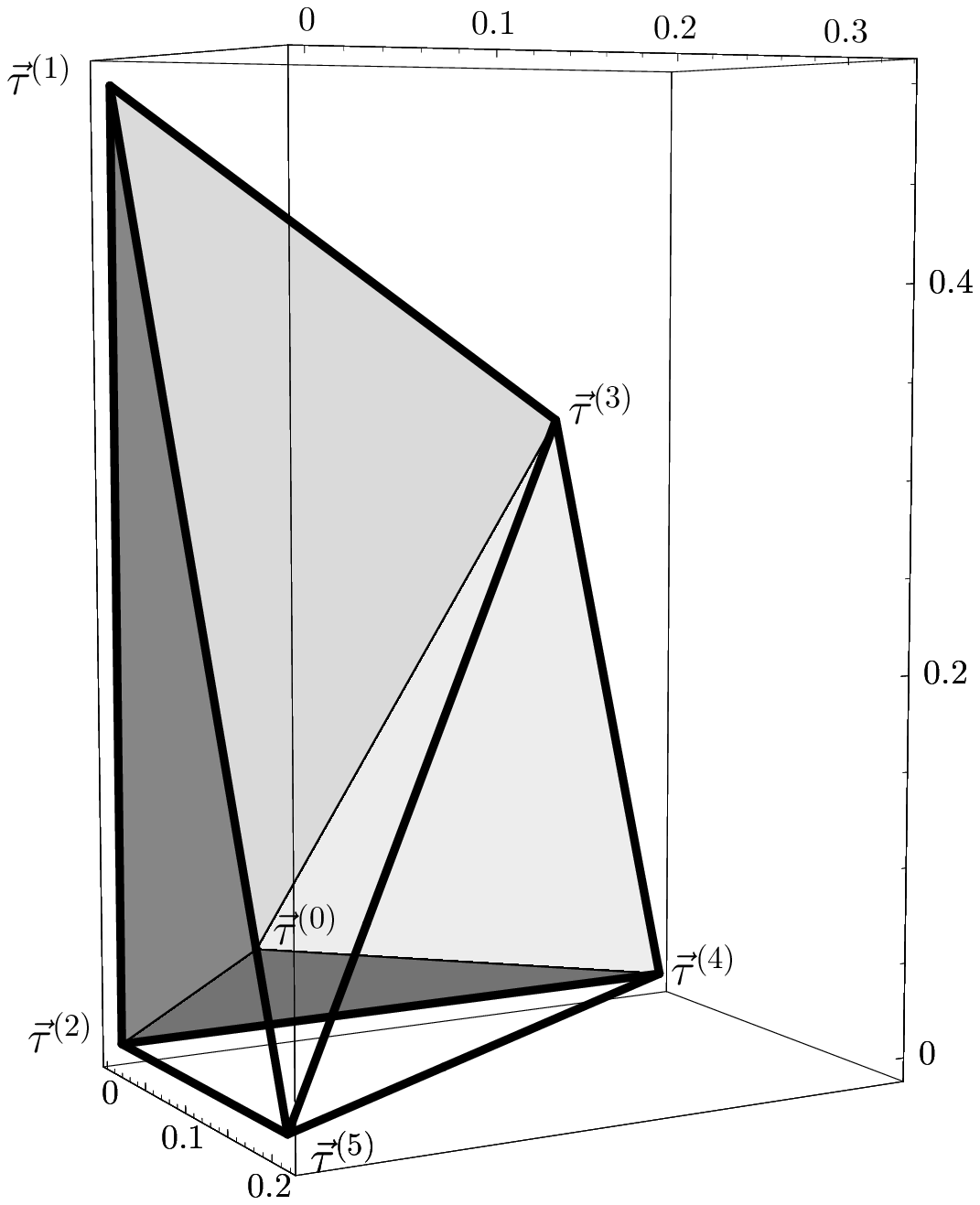,width=8.5cm}
\caption{\label{figppt}The set of symmetric PPT states $\sigma$
(thick wired object) parameterized by the three coordinates
$\lambda_i=\tr{\sigma P_i}$, plotted for $d=3$. The solid object
inside corresponds to the set of separable states. The {\it
universal activators} lie on the plane $\{\vec{\tau}^{(3)},
\vec{\tau}^{(4)},\vec{\tau}^{(5)}\}$ and contain an arbitrary
small amount of entanglement near the line $\{\vec{\tau}^{(3)},
\vec{\tau}^{(4)}\}$.}
\end{figure}

The set $\cal P$ corresponding to normalized operators with a
positive partial transpose can easily be obtained by observing
that the symmetry group of the partially transposed operators
$\sigma^{T_A}$  in (\ref{UUUA}) is equal to the group of unitaries
$W$ when interchanging the systems $1\leftrightarrow 2$. The
respective minimal projectors $\{Q_i\}$ can therefore be obtained
from the projectors $\{P_i\}$ simply by relabeling the systems
$1\leftrightarrow 2$, and the $k$-th coordinate of an extreme
point $\vec{p}^{(i)}$ of $\cal P$ is thus given by
\begin{equation}\label{pi}
  p_k^{(i)}=\tr{Q_i^{T_A}P_k}/\tr{Q_i}.
\end{equation}
Hence, $\cal P$ is again a tetrahedron and Eq. (\ref{pi}) leads to
the extreme points:
$$\ba{lcl}{\vec{p}^{(1)}=\Big(\frac{d-1}{2d},\frac{-d-1}{2d},\frac{1-d^2}{2d}\Big) &,& \vec{p}^{(3)}=\Big(\frac{-1}{2d},\frac{-1}{2d},\frac{d+1}{2d}\Big),\\
\vec{p}^{(2)}=\Big(\frac{1-d}{2d},\frac{1+d}{2d},\frac{-(d-1)^2}{2d}\Big)
&,&
\vec{p}^{(4)}=\Big(\frac{1}{2d},\frac{1}{2d},\frac{d-1}{2d}\Big).}
$$

Straight forward linear algebra now allows us to compute the
extreme points $\{\vec{\tau}^{(i)}\}$ of the intersection ${\cal
S}\cap{\cal P}$ corresponding to the set of symmetric PPT states,
which is shown in Fig.\ref{figppt}:
$$ \ba{lcl}{
\vec{\tau}^{(1)}=\Big(0,0,\frac12 \Big) &,&
\vec{\tau}^{(2)}=\Big(0,0,0\Big),\\\vec{\tau}^{(3)}=\Big(\frac{1}{2d},\frac{1}{2d},\frac{d-1}{2d}\Big)
&,&  \vec{\tau}^{(4)}=\Big(0,\frac1d,0\Big),\\
\vec{\tau}^{(5)}=\Big(\frac{1}{d+2},0,0\Big).}$$

The PPT state space given by the convex hull of these points can
be divided into three parts: separable states and activating
respectively not activating bound entangled states.

\subsection{Separable states}

A vector $\vec{\lambda}\in{\cal S}\cap{\cal P}$ corresponds to a
separable symmetric state iff we can find any (not necessarily
symmetric) separable state $\rho_{sep}$ such that
$\lambda_i=\tr{P_i \rho_{sep}}$ (cf.\cite{Sym}).

A special case of a symmetric separable state is of course a
tensor product of a Werner and an Isotropic state
$\sigma=\rho(\alpha)_1\otimes\omega(f)_2$, where both are
separable. In fact, if we choose these two states lying on the
separable boundaries, i.e., $\alpha\in \{-d,1\}$ and
$f\in\{0,1/d\}$, we retrieve the extreme points $\vec{\tau}^{(1)},
\ldots, \vec{\tau}^{(4)}$.

Another point $\vec{\tau}^{(0)}=(\frac1{2d},0,\frac{d-2}{4d})$,
that will also turn out to be an extreme point of the set of
separable symmetric states, is obtained from the product state:
\begin{equation}\label{phipsi}
\rho_{sep}=|\Phi^+\rangle\langle\Phi^+|_A\otimes
|\Psi^-\rangle\langle\Psi^-|_B ,
\end{equation}
where
$|\Phi^+\rangle=\frac{1}{\sqrt{2}}\big(|11\rangle+|22\rangle\big)$
and
$|\Psi^-\rangle=\frac{1}{\sqrt{2}}\big(|12\rangle-|21\rangle\big)$
are two-dimensional maximally entangled states.

The convex hull of the points $\vec{\tau}^{(0)}, \ldots,
\vec{\tau}^{(4)}$ (see Fig.\ref{figppt}) already covers the entire
separable region as we are going to show in the following that the
complement of this
 polytope within ${\cal S}\cap{\cal P}$ corresponds to bound
entangled states.

\subsection{Bound entangled and activating}

The equation
$f\big(\rho(\alpha);\sigma(\vec{\lambda})\big)=\frac1d $ written
out as $ \sum_{i} c_i(\alpha) \lambda_i = 0 $, with
$$ c_i(\alpha) = \frac{\tr{(\rho(\alpha)\otimes
P_i)(\mathbb{P}_{A_{0,1}} \otimes\mathbb{P}_{B_{0,1}} \otimes(d
\mathbb{P}-{\bf 1} )_2)}}{\tr{P_i}}$$ is linear in $\lambda_i$ and
thus defines a plane separating symmetric states activating
$\rho(\alpha)$ from states apparently not activating it. The task
is now to construct this separating plane depending on the
parameter $\alpha$.

As we have already used above, the points $\vec{\tau}^{(3)},
\vec{\tau}^{(4)}$ correspond to product states of the form
$\sigma=\rho(\alpha)_1\otimes\omega(\frac1d)_2$ for which
$f\big(\rho(\alpha);\sigma\big)=\frac1d$ obviously holds for any
Werner state $\rho(\alpha)$. Thus $\vec{\tau}^{(3)},
\vec{\tau}^{(4)}$ are two fix points of the separating plane with
respect to a variation of $\alpha$ and we need to know only one
more point. For this purpose we consider the line
$\vec{\lambda}(t)=t \vec{\tau}^{(5)} + (1-t) \vec{\tau}^{(1)}$.
Solving the equation
$f\Big(\rho(\alpha);\sigma\big(\vec{\lambda}(t)\big)\Big)=\frac1d$
yields the required third point with
\begin{equation}\label{t}
t=\frac{2+d}{2\alpha+d}.
\end{equation}
This is obviously a strict monotone function in $\alpha$ and it
leads to the following properties of the separating plane:
\begin{enumerate}
  \item For $\alpha=\frac{d}2$ corresponding to the boundary
  Werner state which is not 1-distillable, Eq. (\ref{t})
  leads to $\vec{\lambda}(t)=\vec{\tau}^{(0)}$, showing that
  $\vec{\tau}^{(0)}$ indeed lies on the boundary of the set of
  separable states. That is, any PPT state in front of the plane $\{\vec{\tau}^{(0)},\vec{\tau}^{(3)},
\vec{\tau}^{(4)}\}$ must be bound entangled since it activates at
least $\rho(\alpha=\frac{d}2)$.
  \item In the limit $\alpha\rightarrow 1$, i.e., $\rho(\alpha)$
  becoming less and less entangled, $\vec{\lambda}(t)$ approaches
  $\vec{\tau}^{(5)}$. However, for any $\alpha =1+\varepsilon, \varepsilon > 0 $ the polytope $\{\vec{\tau}^{(0)},\vec{\tau}^{(3)},
\vec{\tau}^{(4)},\vec{\lambda}(t)\}$ has a nonempty interior
corresponding to PPT bound entangled states capable of activating
any $\rho(\alpha)$ with $\alpha>1+\epsilon$.
  \item Except for the line $\{\vec{\tau}^{(3)},
\vec{\tau}^{(4)}\}$ all PPT states on the plane
$\{\vec{\tau}^{(3)}, \vec{\tau}^{(4)},\vec{\tau}^{(5)}\}$ lie on
the activating side of the separating plane for any $\alpha
>1$. The corresponding symmetric states can thus be considered to
be {\it universal activators} in the sense that they activate any
entangled Werner state and therefore any NPPT state.
\end{enumerate}

The set of bound entangled {\it universal activators} contains
states arbitrary close to the line $\{\vec{\tau}^{(3)},
\vec{\tau}^{(4)}\}$ which in turn corresponds to separable states.
By the continuity properties of the entanglement measures {\it
Entanglement of Formation} \cite{EoF} and {\it Relative Entropy of
Entanglement} \cite{RelEnt} this geometric vicinity, however,
translates directly to the proposition that these states contain
an arbitrary small amount of entanglement.

\subsection{Bound entangled and not activating}

In order to complete the classification of the symmetric states
introduced in Eq. (\ref{UUUA}) we have still to determine the
entanglement properties of the states corresponding to the
tetrahedron
$\{\vec{\tau}^{(0)},\vec{\tau}^{(2)},\vec{\tau}^{(4)},\vec{\tau}^{(5)}\}$.
The plane separating this set from the separable states derived
above is characterized by a linear operator $W$ via
$\tr{W\sigma(\vec{\lambda})}=0$, where
\begin{equation}\label{W}
W=\big({\bf 1}-\mathbb{F}\big)_1\otimes\big({\bf
1}-\frac{d}2\mathbb{P}\big)_2.
\end{equation}
However, this operator is an {\it entanglement witness}
(cf.\cite{optwit}), meaning that $\tr{W\rho}\geq 0$ holds for any
separable state $\rho$. In order to see this property we have just
to utilize the results from \cite{IHa,IBM}, where it was shown
that ${\bf 1}-\frac{d}2\mathbb{P}$ has a positive expectation
value on any pure state of Schmidt rank two. Since the
antisymmetric projector $\frac12\big({\bf 1}-\mathbb{F}\big)$ is a
sum of Schmidt rank two states, $\langle\phi_A\otimes\psi_B |W|
\phi_A\otimes\psi_B\rangle$ is a sum of such positive expectations
for any pure product state $| \phi_A\otimes\psi_B\rangle$. Hence,
$\tr{W\rho}\geq 0$ is indeed fulfilled by any separable state
implying that the tetrahedron under discussion corresponds to
bound entangled symmetric states which are however not activating
(with respect to the considered protocol).

\section{Conclusion}
We investigated the entanglement properties of a new abelian set
of symmetric states with regard to the activation of NPPT
distillation. The set contains PPT bound entangled states of full
rank providing a universal resource for the activation of any NPPT
state (in any finite dimension). Some of these {\it universal
activators} lie arbitraryly close to separable states such that
the activation process turns out to require only an infinitesimal
amount of entanglement. This indicates that the difference between
1-distillable and $n$-undistillable or even bound entangled NPPT
states is very subtle. Moreover, the above results show that even
weakly entangled bound entangled PPT states can be useful for some
quantum information processing purposes \cite{BEtele}.

As the problem discussed in this paper is primarily a feasibility
problem, we have at this stage not asked for the obtained rates.
In fact, one could for instance easily improve the probability of
success for the used activation protocol by a factor $d^2$ by
measuring in a basis of maximally entangled states and retaining
the state whenever the measurement outcomes coincide. An
interesting question going one step further and requiring
knowledge about rates is, whether there is the possibility of {\it
self-activation} after some initial activation with a limited
resource took place. In other words, is it possible to yield
asymptotically more entanglement from distillation than is needed
for the preparation of the activator states?

\section*{Acknowledgement}
The authors would like to thank R.F. Werner and T. Eggeling for
stimulating discussions and in particular M. Lewenstein for
bringing the witness property of $W$ to their attention. Funding
by the European Union project EQUIP (contract IST-1999-11053) and
financial support from the DFG (Bonn) is gratefully acknowledged.


\begin{references}
\bibitem{KLC} It was recently shown, however,  that there exist
three-partite NPPT states, with the property that two copies can
neither be distilled nor activated \cite{KLC2}.
\bibitem{BEtele} The first example showing that bound entangled
PPT states can indeed be useful as an additional resource for
teleportation protocols was given in Ref. \cite{BEtele2}.
\end{references}
\end{document}